\documentclass[final,1p,times]{elsarticle}

\usepackage[utf8]{inputenc}
\usepackage{graphicx}
\usepackage{amssymb}
\usepackage{amsmath}
\usepackage{booktabs}
\usepackage{hyperref}
\usepackage{subcaption}
\usepackage[usenames,dvipsnames]{xcolor}
\usepackage{tkz-kiviat} 
\usetikzlibrary{arrows}
\usepackage{adjustbox}

\usepackage{lineno}

\biboptions{comma,square,sort}
\makeatletter
\providecommand{\doi}[1]{%
	\begingroup
	\let\bibinfo\@secondoftwo
	\urlstyle{rm}%
	\href{http://dx.doi.org/#1}{%
		doi:\discretionary{}{}{}%
		\nolinkurl{#1}%
	}%
	\endgroup
}
\makeatother

\makeatletter
\def\ps@pprintTitle{%
	\def\@oddfoot{}%
	\let\@evenfoot\@oddfoot}
\makeatother

\journal{}

\begin{document}

\begin{frontmatter}

\title{Goals and Measures for Analyzing Power Consumption Data in Manufacturing Enterprises}

\author[se]{Sören Henning\corref{cor1}}
\ead{soeren.henning@email.uni-kiel.de}

\author[se]{Wilhelm Hasselbring}
\ead{hasselbring@email.uni-kiel.de}

\author[kn]{Heinz Burmester}
\ead{heinz.burmester@kieler-nachrichten.de}

\author[ibak]{Armin Möbius}
\ead{a.moebius@ibak.de}

\author[wobe]{Maik Wojcieszak}
\ead{mw@wobe-systems.com}

\address[se]{Software Engineering Group, Kiel University, 24098 Kiel, Germany}
\address[kn]{Kieler Zeitung GmbH \& Co. Offsetdruck KG, Radewisch 2, 24145 Kiel, Germany}
\address[ibak]{IBAK Helmut Hunger GmbH \& Co. KG, Wehdenweg 122, 24148 Kiel, Germany}
\address[wobe]{wobe-systems GmbH, Edisonstraße 3, 24145 Kiel, Germany}

\cortext[cor1]{Corresponding author}

\begin{abstract}

The Internet of Things adoption in the manufacturing industry allows enterprises to monitor their electrical power consumption in real time and at machine level. In this paper, we follow up on such emerging opportunities for data acquisition and show that analyzing power consumption in manufacturing enterprises can serve a variety of purposes. Apart from the prevalent goal of reducing overall power consumption for economical and ecological reasons, such data can, for example, be used to improve production processes.

Based on a literature review and expert interviews, we discuss how analyzing power consumption data can serve the goals reporting, optimization, fault detection, and predictive maintenance. 
To tackle these goals, we propose to implement the measures real-time data processing, multi-level monitoring, temporal aggregation, correlation, anomaly detection, forecasting, visualization, and alerting in software.

We transfer our findings to two manufacturing enterprises and show how the presented goals reflect in these enterprises.
In a pilot implementation of a power consumption analytics platform, we show how our proposed measures can be implemented with a microservice-based architecture, stream processing techniques, and the fog computing paradigm.
We provide the implementations as open source as well as a public demo allowing to reproduce and extend our research.

\end{abstract}


\end{frontmatter}

\section{Introduction}

The immense electrical power consumption of the manufacturing industry \cite{IEA2019} is a considerable cost factor for manufacturing enterprises and a serious problem for environment and society. Corporate values, public relations, energy-related costs, and legal requirements are therefore leading to an increasing energy awareness in enterprises \cite{Shrouf2017}. %
At the same time, trends toward the Industrial Internet of Things, Industry 4.0, smart manufacturing, and cyber-physical production systems allow to collect energy data in real time and at machine level, from smart meters or machine-integrated sensors \cite{Shrouf2014a, Mohamed2019}.
Furthermore, research on big data provides methods and technologies to analyze data of huge volume and high velocity, as it is the case with power consumption data \cite{Sequeira2014, Zhang2018}.
However, even though research suggests a variety of goals and measures for analyzing power consumption data, the full potential of available data is rarely exploited \cite{Shrouf2015, Bunse2011, Cooremans2019}.

To summarize our contribution:

\begin{enumerate}
	
	\item We present the findings of a literature review and expert interviews regarding goals and measures for analyzing power consumption data. In Section~\ref{sec:Goals}, we describe categories of such goals. Building upon these goals, we propose a set of software-based measures in Section~\ref{sec:Measures} that serve these goals. Furthermore, we provide a mapping of goals and measures by rating the impact of measures on goals.%
	
	\item We transfer our findings to two enterprises of the manufacturing industry. Both enterprises are project partners of wobe-systems and Kiel University in the Titan project on Industrial DevOps \cite{Hasselbring2019}.
	After a brief overview of energy monitoring in these enterprises in Section~\ref{sec:Enterprises}, we discuss in Section~\ref{sec:Goals-Examples} how the enterprises expect to benefit from analyzing power consumption data. For this purpose, we present specific examples for each goal category.
	
	\item Following on from these use cases, we show with a pilot implementation in Section~\ref{sec:Implementation} how our proposed measures can be implemented in a coherent analytics platform. We provide the implementations as open source as well as a public demo allowing to reproduce and extend our research.
	
\end{enumerate}

Finally, we discuss related work in Section~\ref{sec:Related}, before concluding this paper in Section~\ref{sec:Conclusions}.

\section{Goals for Analyzing Power Consumption Data}\label{sec:Goals}

In this section, we describe how analyzing power consumption data serves the goals of reporting, optimization, fault detection, and predictive maintenance.

\subsection{Reporting}\label{sec:Goals:Reporting}

Reports of analyzed power consumption data assist an enterprise in understanding its energy consumption \cite{Miragliotta2013, Vikhorev2013, Shrouf2017}.
It thus serves as a necessary means for implementing energy management, which requires to reveal all energy consumptions within the enterprise \cite{Fiedler2012}.
Reports must therefore provide insights into which devices, machines, and enterprise departments use how much power and during which times this power is consumed. Combined with information about the production processes, reports can thus be used to identify which processes consume how much power \cite{Herrmann2009}. In this way, measures for optimizing energy consumption can be evaluated and saving potentials can be identified (see also the following Section~\ref{sec:goal-optimization}) \cite{Bunse2011}.

Comprehensive reporting is required in particular for an ISO~50001 \cite{ISO50001} certification. The ISO~50001 standard specifies requirements for organizations and businesses for establishing, implementing, and improving an energy management system. It describes a systematic approach to support organizations in continuously improving their energy efficiency. In order to be certified to use an energy management system in compliance with ISO~50001, enterprises commission accredited certification bodies to perform regular independent audits \cite{Jovanovic2016}. These certifications are usually not required by law, but serve as evidence that a company is making efforts to save energy. For example, in Germany ISO~50001 certification is a prerequisite for manufacturing enterprises with high power consumption to reduce regulatory charges (e.g., reducing the EEG surcharge \cite{BAFA2018}).
Furthermore, the ISO~50001 standard requires that reports on the enterprise's energy consumption are available for customers, stakeholders, employees, and management.

\subsection{Optimization}\label{sec:goal-optimization}

The most prevalent optimization in the context of power consumption is reducing the overall consumption \cite{Bunse2011, Miragliotta2013, Shrouf2014b, Schulze2016, Shrouf2017}.
Besides the evident economic motivations, there are also ecological reasons for making production processes more sustainable.
To achieve this, a first step is to identify energy-inefficient machines and devices. This knowledge can then be used to replace them with more energy-efficient ones or retrofitting them accordingly. Furthermore, time periods should be detected in which devices consume energy, although it would not be necessary. Typical examples of unnecessary energy consumption are keeping machines in standby mode or lighting workplaces outside of working hours, but also less apparent saving potential could be discovered.

Another optimization aspect is the reduction of peak loads \cite{Herrmann2009, Vikhorev2013, Shrouf2015}. In addition to the basic price, which is fixed per month, and the price per kilowatt hour, large-scale power consumers such as manufacturing enterprises often have to pay a demand rate. The demand rate depends on the maximum demand that occurs within a billing period. In this way, grid operators expect to have a load as uniformly as possible in the electricity grid \cite{Albadi2008}.
Demand peaks are therefore disproportionately more expensive for the customer. Thus, an optimization should aim to achieve a power consumption as constant as possible, i.e., to distribute the demand evenly over time (peak shaving).
In order to achieve this, it is necessary to identify periods during which relatively much power is demanded. Likewise, it is important to discover which consumers are responsible for the demand and to what extent \cite{Herrmann2009}.
This includes, on the one hand, the identification of large consumers in general but, on the other hand, also demand fluctuations of individual devices. Based on this information, production processes can be modified such that, for instance, multiple machines with a high inrush current are not started at the same time.
Reducing the overall energy consumption is highly related to reducing peak loads. If measures are taken to replace devices, this has an effect on both optimization goals. For example, if devices that are unnecessarily operated standby during load peaks are turned off during these periods, not only demand peaks are reduced, but also the enterprise's power consumption in total.

In order to reduce both total consumption and load peaks, it can be expedient to optimize the operation of machines in production individually \cite{Shrouf2014b}. 

\subsection{Fault Detection}

If the power consumption of machines deviates from its normal behavior, this can be an indicator for a fault such as a mechanical defect or faulty operation \cite{Quiroz2018}. Analyzing the power consumption can therefore be used to automatically detect such faults and to react accordingly \cite{Vijayaraghavan2010, Mohamed2019}. A typical case of anomalous power consumption is a strong increase, for example, when a defect occurs suddenly. A decrease of power consumption can also be such an indicator as parts of a machine may no longer be operated due to a defect. Less noticeable is a slight deviation over a longer period of time, for example, if several minor defects occur over time. Detecting deviations or a long-term trend in regularly fluctuating power consumption is even more challenging (see our industrial case study in Section~\ref{sec:Goals-Examples:Fault-Detection}).

\subsection{Predictive Maintenance}

With regular, time-based maintenance intervals, machines and devices are often maintained even though there is no actual need for it. This means that components and operating materials are replaced since their expected operating time expires, although they are still functioning and could actually continue operating. Predictive maintenance is an approach that aims for performing maintenance actions only if it would otherwise results in defects or limitations in performance or quality \cite{Yan2017}. The difficulty is therefore to decide when maintenance is really necessary. For this purpose, sensor data of the machine and its environment are collected and automatically analyzed \cite{Yan2017}. Power consumption can be one of those correlating factors \cite{Shrouf2014a, Mohamed2019}.

While fault detection aims to detect errors after they occurred, predictive maintenance refers to the detection of errors before they occur. Nevertheless, predictive maintenance is closely related to fault detection as occurring faults often cause further faults. Therefore, early fault detection may allow future faults to be detected and appropriate preventive measures to be taken.

\section{Measures for Analyzing Power Consumption Data}\label{sec:Measures}

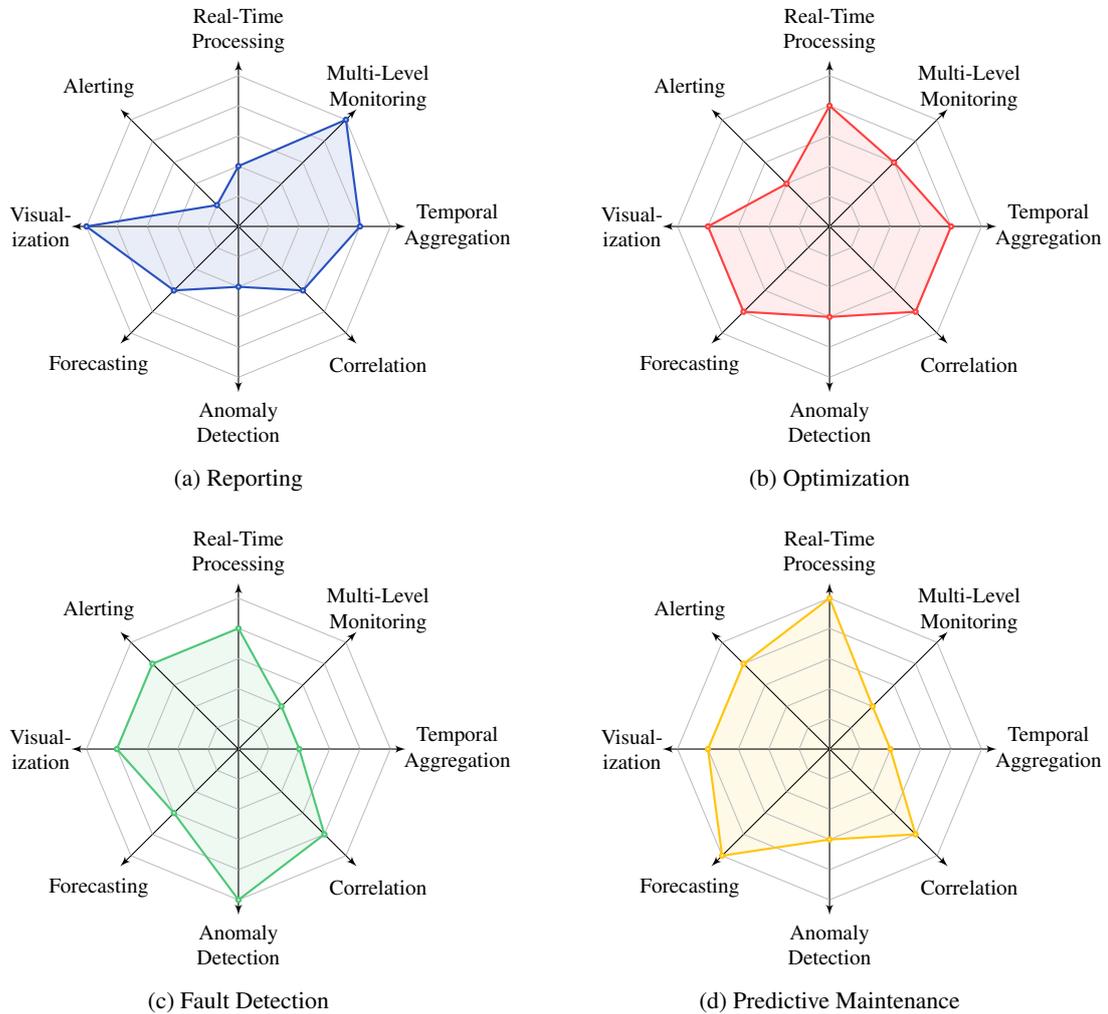
\begin{figure*}[t]
	\newcommand{\chart}{
		\footnotesize
		\tkzKiviatDiagram[
			scale=0.4,
			label distance=.2cm,
			label space = 1.5,
			radial  = 8,
			gap     = 1,  
			lattice = 5,
			yscale  = 1,
			xscale  =-1,
			rotate  = 90
		]{Real-Time Processing,Multi-Level Monitoring, \hspace{2em}\parbox{5em}{\centering Temporal Aggregation}, Correlation, Anomaly Detection, Forecasting, \parbox{5em}{\centering Visual\-ization}\hspace{2em}, Alerting}
	}
	\definecolor{ceruleanblue}{rgb}{0.16, 0.32, 0.75}
	\definecolor{coralred}{rgb}{1.0, 0.25, 0.25}
	\definecolor{emerald}{rgb}{0.31, 0.78, 0.47}
	\definecolor{mikadoyellow}{rgb}{1.0, 0.77, 0.05}
	\definecolor{niceblue}{named}{ceruleanblue}
	\definecolor{nicered}{named}{coralred}
	\definecolor{nicegreen}{named}{emerald}
	\definecolor{niceorange}{named}{mikadoyellow}

\begin{adjustbox}{minipage=[t]{1.13\textwidth},center}
	\begin{subfigure}[b]{0.49\textwidth}%
		\centering
		\begin{tikzpicture}
			\chart
			\tkzKiviatLine[thick,color=niceblue,mark=*,
			fill=niceblue!20,opacity=.5](2,5,4,3,2,3,5,1)
		\end{tikzpicture}
		\caption{Reporting}
		\label{fig:reporting}
	\end{subfigure}
	\hfill%
	\begin{subfigure}[b]{0.49\textwidth}
		\centering
		\begin{tikzpicture}
			\chart
			\tkzKiviatLine[thick,color=nicered,mark=*,
			fill=nicered!20,opacity=.5](4,3,4,4,3,4,4,2)
		\end{tikzpicture}
		\caption{Optimization}
		\label{fig:optimization}
	\end{subfigure}

	\vspace{1em}
	
	\begin{subfigure}[b]{0.49\textwidth}
		\centering
		\begin{tikzpicture}
			\chart
			\tkzKiviatLine[thick,color=nicegreen,mark=*,
			fill=nicegreen!20,opacity=.5](4,2,2,4,5,3,4,4)
		\end{tikzpicture}
		\caption{Fault Detection}
	\end{subfigure}
	\hfill %
	\begin{subfigure}[b]{0.49\textwidth}
		\centering
		\begin{tikzpicture}
			\chart
			\tkzKiviatLine[thick,color=niceorange,mark=*,
			fill=niceorange!20,opacity=.5](5,2,2,4,3,5,4,4)
		\end{tikzpicture}
		\caption{Predictive Maintenance}
	\end{subfigure}
\end{adjustbox}

	\caption{Impact rating of the suggested measures for the four goals presented in Section~\ref{sec:Goals} based on expert interviews. The larger its distance from the radar chart's center is, the higher a measure's impact was weighted on the corresponding goal.}
	\label{fig:goals-measures-charts}
\end{figure*}

In this section, we discuss software-based measures for analyzing power consumption data that support in achieving the goals defined in the previous section. We suggest the following measures: real-time data processing, multi-level monitoring, temporal aggregation, correlation, anomaly detection, forecasting, visualization, and alerting. Different use cases weight goals differently and measures vary in their importance for the individual goals. We therefore rate the impact of each measure on each goal and visualize these impacts on radar charts shown in Figure~\ref{fig:goals-measures-charts}. In the following, we describe each of the suggested measures in detail.

\subsection{Near Real-time Data Processing}

Near real-time (also referred to as online) data processing describes approaches, where data are immediately processed after their recording. It contrasts batch (also referred to as offline) processing, which first collects recorded data and then processes all the collected data only at certain times. Whereas near real-time data processing is usually more difficult to design and implement than batch processing, it yields immediate results and, thus, allows to react immediately on these results.

Data processing in near real-time supports primarily the goals optimization, fault detection, and predictive maintenance \cite{Vijayaraghavan2010, Shrouf2015}.
Power consumption can be efficiently optimized if the effectiveness of energy-saving actions are evaluated immediately.
The sooner a fault is detected and reported, the faster it can be reacted to the fault and, therefore, the more valuable its detection is.
Predictive maintenance requires processing monitoring data in real time as otherwise the time for maintenance may be determined after the maintenance should have already been performed \cite{Sahal2020}.
Although a real-time overview of the enterprise's energy usage at any time is not required for ISO~50001 audits, it assists in reporting the power consumption, for example, to the management \cite{Miragliotta2013}.

\subsection{Multi-Level Monitoring}\label{sec:MultiLevelMonitoring}

Individual goals require data to be collected at different levels \cite{Vikhorev2013, Shrouf2015,Kanchiralla2020}. Whereas, for example, the effect of overall power consumption optimizations can be evaluated with data of the overall power consumption, detecting defects in machines requires to acquire data at machine-level. Moreover, different stakeholders are often interested in power consumption reports of different granularity \cite{Shrouf2017}.

We suggest to organize power consumers in a hierarchical model, where groups of devices and machines are further grouped into larger groups \cite{Henning2019b}. Multiple such models have to be maintained in parallel. For example, it is reasonable to organize devices by their type (e.g., all air compressors), but also to organize them by their physical location (e.g., a certain shop floor).
Such a hierarchical breakdown in particular supports reporting as it offers insights at which times which consumers or groups consume most power.

Besides monitoring groups of consumers, for example, via sub-distribution units, data for groups can also be obtained by aggregating the consumption of all its partial consumers. In particular, this is necessary for devices which, for reasons of redundancy, have more than one power supply. Here, the overall machine's power consumption is usually more important than the power consumption of the individual power supplies.
Comparing the power consumption monitored by sub-distribution units with aggregated data of all known sub-consumers may reveal consumptions, which were unknown so far.

\subsection{Temporal Aggregation}\label{sec:TemporalAggregation}

In addition to aggregating the power consumption of multiple consumers to larger groups, it is often required to aggregate multiple measurements of the same consumer over time \cite{Shrouf2015}. Summarizing multiple data points to one data point primarily serves for: (1) reducing the number of data points for storage and (2) simplifying data analysis by providing a more abstract view on the data. Therefore, temporal aggregation supports humans in comprehending the data and, thus, reporting as well as manual identifying optimization potentials.
Further, data processing by automatic processes for optimization, fault detection, and predictive maintenance may benefit from aggregated data. 
We distinguish two different kinds of temporal aggregation as described in the following.

\paragraph{Aggregating Tumbling Windows} %
The first kind is to collect and aggregate all measurements in consecutive, non-overlapping, fixed-sized time windows (tumbling windows \cite{Li2005}).
An appropriate size for such windows is, for example, 5~minutes so that every 5~minutes a new aggregation result is computed representing the average, minimal, and maximal power consumption over the previous 5~minutes. The number of data points can thus be massively reduced, which is required for several forms of storing, analyzing, and visualizing data. We suggest to perform multiple such aggregation (e.g., for time windows of size 1~minute, 5~minutes, and 1~hour) and store their aggregation results for different durations.
This allows to store more recent (and more interesting) data with more detail than data from the previous months or years. %

\paragraph{Aggregating Temporal Attributes}
The second kind of temporal aggregation is to aggregate all data points having the same temporal attribute such as day of week or hour of day. The set of aggregated data points allows to model or identify seasonality. For example, aggregating all measurements recorded at the same day of week allows to show the average power consumption course over a week. %
Likewise, aggregating based on the hour of the day allows to obtain the average course of a day.

\subsection{Correlation}\label{sec:Correlation}

Analyzing energy data often yield significantly better results if, in addition to recorded power consumption, further information is included such as operational and planning data from the production as well as business data \cite{Vijayaraghavan2010,  Shrouf2017}.

Correlating power consumption data with production data supports reporting as it allows to understand why the consumption behaves as it does.
Management levels might be interested in a correlation with business data as this allows to report about, for example, the energy costs per produced unit.
In particular, correlation can serve as trigger for optimization, fault detection, and predictive maintenance. If, for example, the power consumption of a machine increases rapidly while also the production speed increases, the increasing power consumption was most likely not caused by a fault. If, however, the production speed remains constant and no other production data justifies the increase, a fault detection could be triggered.

Furthermore, it is reasonable to correlate the power consumption of different consumers. The power consumption of different machines may depend on each other if their production processes are depended \cite{Bischof2018}. Such dependencies of power consumption are interesting for reporting, but also for optimizations, in particular for reducing load peaks.

\subsection{Anomaly Detection}\label{sec:AnomalyDetection}

Anomaly detection (also referred to as outlier detection) describes methods to automatically find unexpected pattern in data \cite{Chandola2009}.
As we typically consider power consumption in a temporal context, anomaly detection refers to detecting times during which the actual power consumption differs from what would be expected.

A common approach for anomaly detection on power consumption time series data is to make a prediction of the future power consumption based on previous data and, then, to compare this predicted value with the actual (monitored or aggregated, see Section~\ref{sec:MultiLevelMonitoring}) value \cite{Chou2014, Liu2018}.
If the actual consumption differs too much from the predicted value, the corresponding value is considered as an anomaly. For this purpose, each data point is assigned an anomaly score, which is computed using a distance function. If this anomaly score exceeds a previously defined threshold, the corresponding data point is considered as an anomaly. It highly depends on the application scenario, which forecast method, anomaly score metric, and anomaly score threshold should be chosen.

Primarily, anomaly detection serves as a measure for the goal of fault detection. Faults in devices, machines, or production processes are deviations from the desired behavior and, thus, anomalous behavior of power consumption may indicate an occurring fault.
Detecting anomalies in power consumption exclusively in relation to time is often not sufficient. The consumption of many devices is subject to external influences such as temperature \cite{Liu2018} and, especially in production environments, the operating times of machines do not follow daily or weekly patterns \cite{Bischof2018}. Correlating power consumption with environmental, operational, and planing data (see Section~\ref{sec:Correlation}) therefore assists in detecting anomalies.

Furthermore, anomaly detection allows to identify potential applications of optimization and predictive maintenance and supports in explaining power consumption behavior in reporting.

\subsection{Forecasting}\label{sec:Forecasting}

In order to predict the future power consumption, a model of the past power consumption is created, explicitly or implicitly, which is projected into the future \cite{Martinez2015}. Common forecast approaches use statistical methods such as ARIMA \cite{Chujai2013} or kernel density estimation \cite{Arora2016}, machine learning methods such as artificial neural networks \cite{Din2017, Zheng2017}, or a combination of both \cite{Chou2014}.
Whereas much research exists on predicting energy consumption for households \cite{Chujai2013, Liu2018}, buildings \cite{Arora2016, Chou2014}, and electricity grids \cite{Din2017, Zheng2017}, approaches on forecasting power consumption of industrial production environments are rare due to their irregular nature \cite{Bischof2018}. Correlating power consumption with environmental, operational, and planning data (see Section~\ref{sec:Correlation}) therefore promises to assist in predicting future power consumption.

Forecasting the power consumption of machines in addition to the overall production environment supports optimization as it allows to detect load peaks before they actually occur. Thus, production operators may take appropriate countermeasures such as replanning production processes.
Making predictions about the future status of the production environment is required for predictive maintenance. Thus, forecasting power consumption enables predictive maintenance based on power consumption data.
As described in Section~\ref{sec:AnomalyDetection}, fault detection based on anomaly detection requires forecasts in order to compare the actual consumption with the expected one. Furthermore, forecasting can be used in reporting as it supports planing and decision making for business and production operation.

\subsection{Visualization}\label{sec:Visualization}

Most approaches for energy analytics platforms and energy management systems include data visualizations \cite{Fiedler2012, Vikhorev2013, Sequeira2014, Zhang2018}. These visualization integrate several of the measures proposed in this chapter and, thus, serve as a link between data analysis and the users of such systems.  Visualizations are often realized as a dashboard, which contains multiple components providing different types of visualization.
Individual components show, for example:
\begin{itemize}
	\item the current status of power consumption as numeric values or gauges \cite{Rist2019, Vikhorev2013}
	\item the evolution of consumption over time in line charts \cite{Vikhorev2013, Sequeira2014, Fiedler2012}
	\item the distribution among subconsumers and categories (also in the course of time) \cite{Vikhorev2013, Masoodian2015}
	\item correlations of individual power consumer \cite{Sequeira2014, Masoodian2017}
	\item particular important values such as the peak load \cite{Vikhorev2013}
	\item detected anomalies \cite{Chou2017}
	\item forecasted power consumption \cite{Singh2018}
\end{itemize}
These dashboards should be dynamic and interactive in the sense that they are updating their visualized data continuously and let users interact with them \cite{Rist2019}.
For example, dashboards may start with a rough outline of the overall production's power consumption but allow users to zoom in and show specific machines and time periods in detail.

First and foremost, dashboards enable reporting on power consumption. Appropriate visualization allows to understand how power consumption is composed, observe changes in power consumption over time, and compare the power consumption of different machines and production processes. Enterprises may provide different dashboards for different stakeholders to only show the information, which is relevant for the corresponding target audience \cite{Shrouf2015}.
Visualizations assist in optimization as they allow to identify optimization potentials and enable operators to check whether optimization actions are effective. Furthermore, interactive visualizations can motivate, trigger, and enable energy saving actions \cite{Rist2019}.
A dashboard may also show information concerning fault detection and predictive maintenance and provide means to verify whether faults and maintenance actions are detected successfully \cite{Shrouf2015}.

As state-of-the-art libraries and framework for data visualization are largely based on web technologies, it is reasonable to implement dashboards as web applications. This has the additional advantage that the visualization is user-friendly accessible since it does not have further requirements on software or hardware infrastructure than a web browser.

\subsection{Alerting}

Industrial production becomes increasingly autonomous \cite{Lasi2014}. Permanently observing a dashboard (see Section~\ref{sec:Visualization}) and waiting for faults or necessary maintenance to be detected can therefore be a tedious work. Instead, it would be convenient to automatically notify production operators when faults are detected or maintenance actions have to be taken \cite{Liu2018}. Depending on their frequency and severity, such notifications and alerts may be sent via email or messenger. For reporting purposes, such notifications may additionally be displayed in a dashboard. Furthermore, operators may be notified if optimization potential is detected, for example, by generating an alert if a load peak is about to occur.

\section{Studied Pilot Cases}\label{sec:Enterprises}

In the following sections, we show how the presented goals apply to two manufacturing enterprises and demonstrate how our suggested measures can be implemented. In this section, we give a brief overview of the two studied enterprises.

\begin{figure*}[t]
	\centering
	\includegraphics[width=\textwidth]{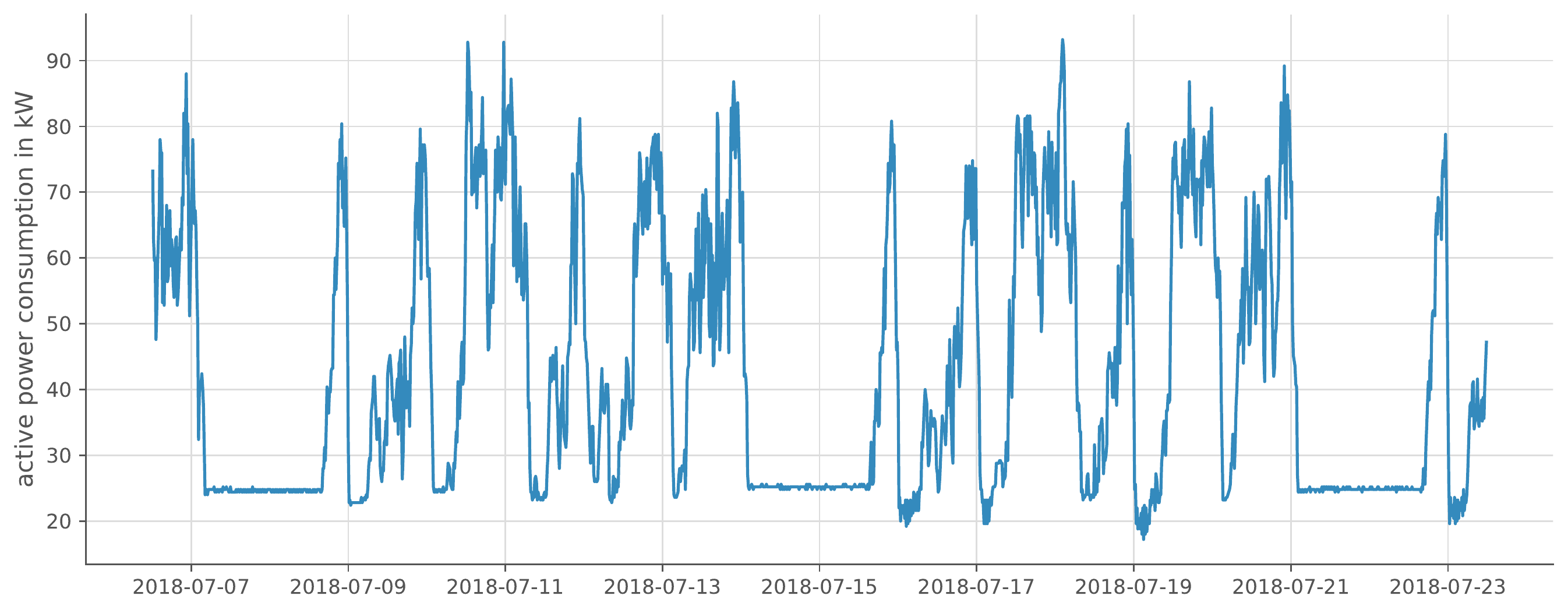}
	\caption{Power consumed for generating compressed air in the newspaper printing company over a period of 2.5 weeks. The curve shows a weekly pattern and reflects the company's operating hours with constant low consumption at the weekend and peak loads at night.}
	\label{fig:compressed-air-two-weeks}
\end{figure*}

The first studied enterprise is a newspaper printing company. It is characterized by high requirements on production speed and the fact that production downtimes are extremely critical. The company has to print and deliver daily newspapers for the next day within a few hours during the night. If newspapers would be printed too late, they are not up to date anymore and could no longer be sold. Production failures would therefore be associated with significant economic damage.
The characteristic production times, with peaks in the nights before working days, are reflected, for example, in the power consumption of the air compressors as depicted in Figure~\ref{fig:compressed-air-two-weeks}.
In addition to daily newspaper printing, the company prints advertising supplements, weekly newspapers, and customer magazines to utilize production capacity.

The second studied enterprise is a manufacturer of optical inspection systems for non-man-size pipelines and wells. This enterprise is characterized by a high vertical range of manufacturing. Thus, its production environment operates a wide range of machines, some of which are largely autonomous, others are primarily user-controlled. Furthermore, the manufacturer operates a rather large data center which runs software for its administration, development, and production. In this paper, we focus on power consumption of the production processes and not on the power consumption of inspection systems themselves.

Both enterprises already have the necessary physical infrastructure to record electrical power consumption in production and query it during operation. Electricity meters already capture the required data with great detail, that is, at machine level and with high frequency. We therefore do not include approaches and techniques for acquiring power consumption data in this paper.
However, both companies do not yet exploit the full potential of the recorded and stored data. Currently, they analyze the data mainly by hand and only at certain times. Much of the information hidden in power consumption data is therefore not revealed yet. The reasons for this cannot be found in missing interest, but in a lack of applicable technologies. Currently, the production operators use software provided by metering device manufactures, for example, to visualize the stored data. However, this software does not meet all requirements. For example, the amount of visualized data is too large, thus making it hard to extract the really important information. Another issue is the integration of different types of electricity meters. Although standardized protocols exist, many metering devices and systems do not apply them.

\section{Goals in the Studied Pilot Cases}\label{sec:Goals-Examples}

In this section, we present how the goals for analyzing power consumption data presented in Section~\ref{sec:Goals} apply to the studied enterprises.

\subsection{Reporting}

Detailed reporting is required in both studied enterprises for ISO 50001 certification. Both companies consider sustainability as an important pillar of their corporate philosophy. ISO 500001 certification allows them to demonstrate their efforts in saving energy to customers and other stakeholders.
Moreover, the certification enables cost savings as it allows these enterprises to reduce regulatory charges as presented in Section~\ref{sec:Goals:Reporting}.
Certification is even essential for the manufacturer of optical inspection systems. Its customers are mainly public authorities, which often require ISO 50001 certification in their calls for tender. 

Reports for ISO 50001 certification are required to justify irregular or increasing power consumption. This is in particular challenging for the newspaper printing company, where power consumption highly depends on the production utilization and external influences. Hence, this company requires to perform complex analyses for their reports, such as correlations with external data from the production and the environment.

\subsection{Optimization}

The motivations for optimizing energy consumption as presented in  Section~\ref{sec:goal-optimization} also apply to both our studied enterprises. This includes reducing the overall power consumption as well as peak loads for ecological and economical reasons. In order to support production operators in making optimizations, continuous insights into the production's power consumption are necessary. Such insights must provide information on how overall consumption is impacted by individual machines, production planning, and external influences, such as the weather. Moreover, it has to allows for continuous review of taken optimization measures.

A potential power saving measure in the newspaper printing company exists in the printing process. The number of newspapers produced per time unit depends directly on the operating speed of the printing presses. To determine an optimal printing speed, several other factors are also taken into account, such as reliability, which decreases when increasing production speed. Monitoring and analyzing the printing presses' power consumption allows to also include energy-related costs when determining the production speed.

\subsection{Fault Detection}\label{sec:Goals-Examples:Fault-Detection}

\begin{figure*}[t]
	\centering
	\includegraphics[width=\textwidth]{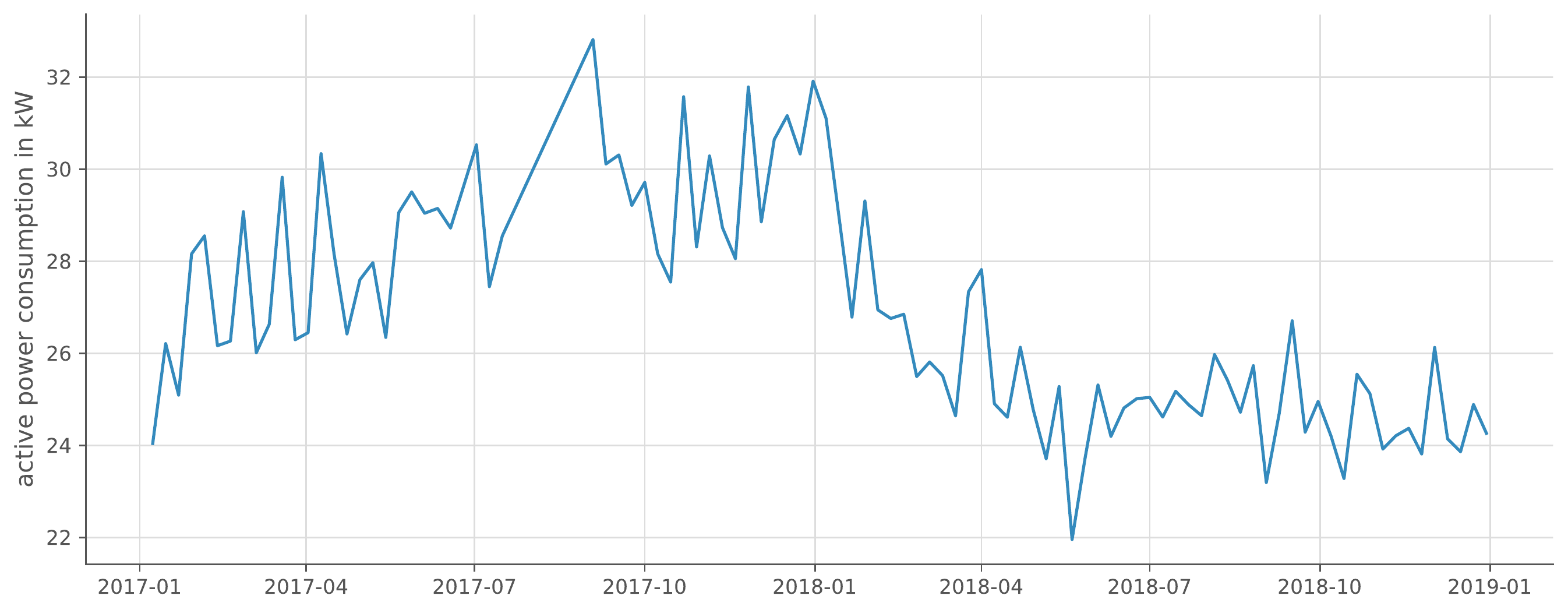}
	\caption{Stand-by power consumption for generating compressed air in the newspaper printing company at weekends over a period of two years.}
	\label{fig:compressed-air-weekends}
\end{figure*}

The central compressed air supply in the newspaper printing company is an example for fault detection using power consumption data. There, an extensive pipe network supplies various areas of the production environment and finally individual machines with compressed air. The compressed air distribution network leaks regularly, causing air to escape. These leaks do not necessarily become apparent directly, but should still be repaired. As leaks result in higher power consumption of the air compressors, power consumption data can provide an instrument for leak detection. However, since power consumption of the air compressors is subject to strong, irregular fluctuations (see Figure~\ref{fig:compressed-air-two-weeks}), an increase in power consumption does not immediately become apparent. This may be solved by considering the power consumption only in idle times, for example, during the weekend. An increase in power consumption over several weekends may thus be an indication of a leak. 
Figure~\ref{fig:compressed-air-weekends} shows the average power consumption between Saturday 12:00 and Sunday 12:00 for each weekend in 2017 and 2018. The course shows a steady increase in 2017 due to leaks in the compressed air supply. In early 2018, the company repaired several leaks, causing a tangible reduction in power consumption.

\subsection{Predictive Maintenance}

An example for predictive maintenance using power consumption are cooling circuits as used in the studied enterprises. Such circulation systems typically include a filter through which a pump pumps coolant to remove impurities. These filters need to be replaced regularly. The electrical power consumption of the pump indicates the resistance within the circulation system and thus how polluted the filter is. Increased power consumption can therefore serve for detecting an upcoming filter change. Lower power consumption can also provide information. It may indicate that not enough coolant is in the circuit (referred to as dry run) and thus coolant needs to be refilled.

\section{Pilot Implementation of Measures}\label{sec:Implementation}

In this section, we show how the measures proposed in Section~\ref{sec:Measures} can be implemented in a software architecture that adopts the microservice architecture pattern, big data stream processing techniques, and fog computing.
In our Titan project on Industrial DevOps \cite{Hasselbring2019}, we develop methods and techniques for integrating Industrial Internet of Things big data. A major emphasis of the project is to make produced data available to various stakeholders in order to facilitate a continuous improvement process. The Titan Control Center\footnote{\url{https://github.com/cau-se/titan-ccp}} is our open source pilot application for integrating, analyzing, and visualizing power consumption data from various sources within industrial production.

The architecture of the Titan Control Center follows the microservice pattern \cite{Newman2015}. It consists of loosely coupled components (microservices) that can be developed, deployed, and scaled independently of each other \cite{Hasselbring2017}. Our architecture features different microservices for different types of data analysis. Individual microservices do not share any state, run in isolated containers \cite{Bernstein2014}, and communicate only via the network. This allows each microservice to use an individual technology stack, for example, to choose the programming language or database system that fits the service's requirements best. In a previous paper \cite{Henning2019a}, we show how these architecture decisions facilitate scalability, extensibility, and fault tolerance of the Titan Control Center.

\begin{figure*}[tb]
	\centering
	\includegraphics[width=\textwidth]{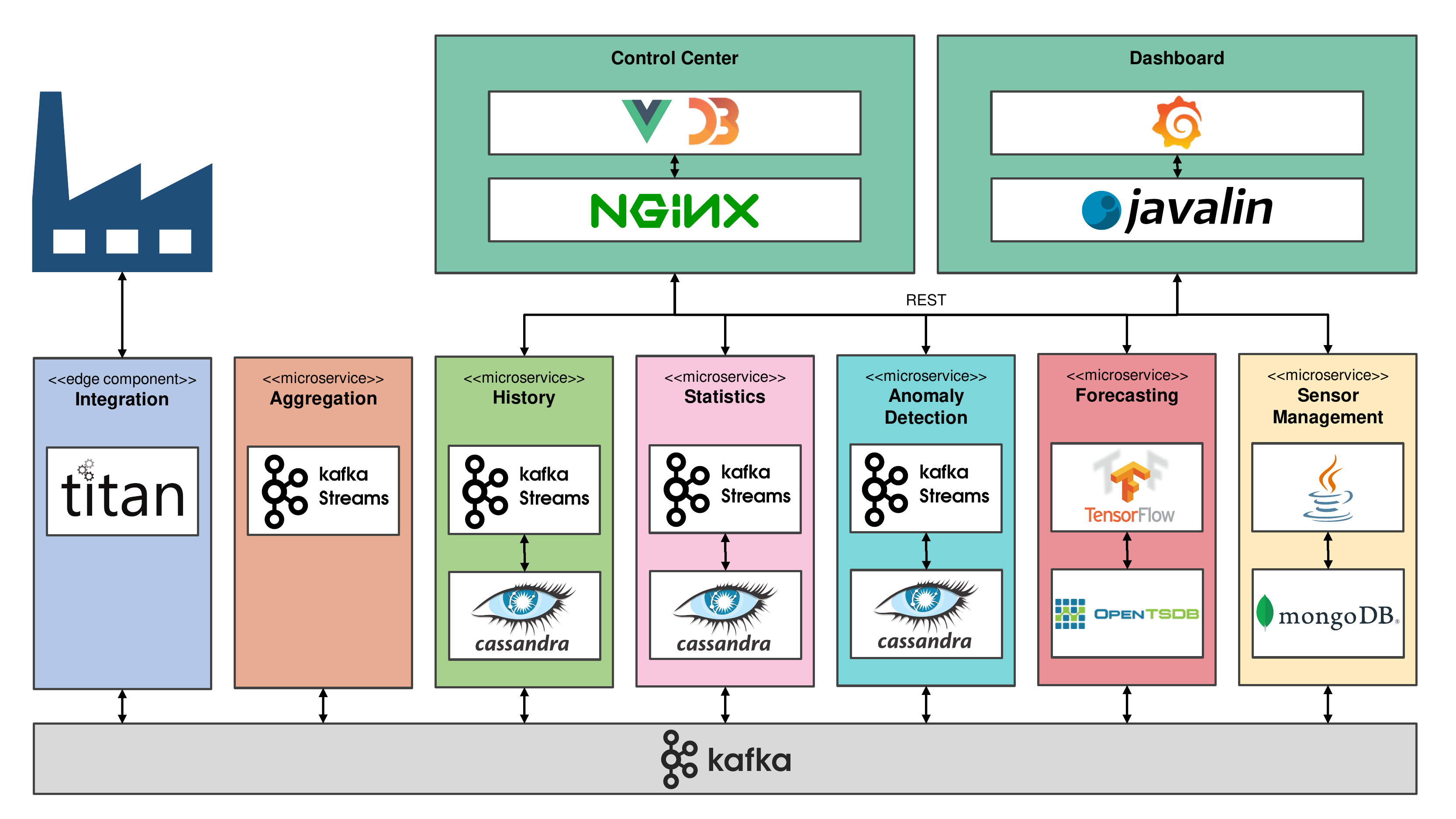}
	\caption{Microservice-based pilot architecture of the Titan Control Center for analyzing electrical power consumption.}
	\label{fig:architecture}
\end{figure*}

Figure~\ref{fig:architecture} shows the Titan Control Center architecture. It contains the microservices Aggregation, History, Statistics, Anomaly Detection, Forecasting, and Sensor Management. In addition to these microservices, our architecture comprises components for data integration, data visualization, and data exchange.

The Titan Control Center is deployed following the concepts of edge and fog computing \cite{GarciaLopez2015, Bonomi2012}.
In particular suited for Internet of Things (IoT) data streams, with these concepts data is preprocessed at the edges of the network (i.e., physically close to the IoT devices), whereas complex data analytics are performed in the cloud \cite{Pfandzelter2019}.
In order to facilitate scalability and fault tolerance, the Titan Control Center microservices for data analysis and storage are deployed in a cloud environment. This can be a public, private, or hybrid cloud, which allows elastic increasing and decreasing of computing resources.
On the other hand, software components for integrating power consumption data into the Titan Control Center are deployed within the production. This includes querying or subscribing to electricity meters, format and unit conversions, filtering, but also aggregations to reduce the amount of data points. We employ our Titan Flow Engine \cite{Hasselbring2019} for this purpose. It allows graphical modeling of data flows in industrial production according to flow-based programming \cite{Morrison2010}.
With the Titan Flow Engine individual processing steps are implemented in so-called \textit{bricks}, which are connected via a graphical user interface to \textit{flows}.
This enables production operators to reconfigure power consumption data flows, for example, to integrate new electricity meters, without having advanced programming skills.

All communication among microservices as well as between the data integration and microservices takes place asynchronously via a messaging system. We use Apache Kafka \cite{Kreps2011} in our pilot implementation. Moreover, the Titan Control Center features two single-page applications that visualize analyzed data and allows for configuring the analyses.

In the following, we present how each measure proposed in Section~\ref{sec:Measures} can be implemented using the Titan Control Center.

\subsection{Near Real-Time Data Processing}

Power consumption data is processed in near real time at all architectural levels of the Titan Control Center.
This start by the ingestion of monitoring data and immediate filter, convert, and aggregate operations in the Titan Flow Engine at the edge. The final integration step is sending the monitoring data to the messaging system.
Following the publish-subscribe pattern, microservices subscribe to this data stream and are notified as soon as new data arrives. In the same way, individual microservices communicate with each other asynchronously.
Apache Kafka as the selected messaging system is proven for high throughput and low latency \cite{Goodhope2012}.
Within microservices, we process data using stream processing techniques \cite{Cugola2012}. This implies that microservices continuously calculate and publish new results as new data arrives. For implementing stream processing architectures in most of the microservices we use Kafka Streams \cite{Sax2018}.
As all computations are performed in near real time, also the visualizations can be updated continuously. Hence, the visualization applications (see Section~\ref{sec:VisualizationImpl}) periodically request new data from the individual services.

\subsection{Multi-Level Monitoring}

The Aggregation microservice \cite{Henning2019b} of the Titan Control Center computes the power consumption for groups of machines by aggregating the power consumption of the individual subconsumers. This microservice subscribes to the stream of power consumption measurements coming from sensors, aggregates these measurement continuously according to configured groups, and publishes the aggregation results via the messaging system as if they were real sensor measurements. In addition to sensor measurements, however, these data can be enriched by summary statistics of the aggregation.

As proposed in Section~\ref{sec:MultiLevelMonitoring}, the Aggregation microservice supports aggregating sensor data in arbitrary nested groups and multiple such nested group structures in parallel.
In one of our studied enterprises, we integrate power consumption data of different kinds of sensors, which provide data in different frequencies. %
An important requirement for the Aggregation service was therefore to support different sampling frequencies. %
Furthermore, besides the focus on scalability throughout the entire Control Center architecture, an important requirement for this microservice is to reliably handle downtimes and out-of-order or late arriving measurements. Therefore, it allows to configure the required trade-off between correctness, aggregation latency, and performance. %

The Sensor Management microservice of the Titan Control Center allows to assign names to sensors and arrange these sensors in nested groups. For this purpose, the Titan Control Center's visualization components provides a corresponding user interface. The Sensor Management service stores these configuration in a MongoDB \cite{MongoDB} database. It publishes changes of group configurations via the messaging system %
such that the Aggregation service (and potentially other services) are notified about these reconfigurations. The Aggregation service is designed in a way that, when receiving reconfigurations, it immediately starts aggregating measurements according to the new groups structure. Further, as aggregations are performed on measurement time and not on processing time, it supports reprocessing historical data.

\subsection{Temporal Aggregation}\label{sec:TemporalAggregationImpl}

Both types of temporal aggregations discussed in Section~\ref{sec:TemporalAggregation} are supported by the Titan Control Center. As both types serve different purposed, they are implemented in individual microservices. Both services subscribe to input streams, which provide monitored power consumption from sensors as well as aggregated power consumption for groups of machines.

\paragraph{Aggregating Tumbling Windows} %
The History microservice receives incoming power consumption measurements and continuously aggregates all data items within consecutive, non-overlap\-ping, fixed-sized windows. The results of these aggregations are stored to an Apache Cassandra \cite{Lakshman2010}
database as well as published for other services. The History service supports aggregations for multiple different window sizes in parallel, allowing to generate time series with different resolutions. To prevent the amount of stored data from becoming too large, time series of different resolutions are assigned different \textit{times to live}. Thus, the Titan Control Center allows, for example, to store raw measurements captured with high frequency for only one day, but aggregated values in minute resolution for years. Window sizes and times to live can be individually configured according to requirements for trackability and availability of storage infrastructure.

\paragraph{Aggregating Temporal Attributes}
The Statistics microservice aggregates power consumption measurements by a temporal attribute (e.g., day of week) to determine an average course of power consumption, for example, per week or per day. These statistics are continuously recomputed, stored in a Cassandra database,
and published for other services, whenever new input data arrives.
In our studied pilot cases we found out that in particular the average consumptions over the day, the week, and the entire year allow to detect pattern in the consumption. Furthermore, aggregating temporal attributes such as the month of the year over one year allows to observe how monthly peak loads evolve over time.

\subsection{Correlation}\label{sec:CorrelationImpl}

\begin{figure*}[t]
	\centering%
	\includegraphics[width=\linewidth]{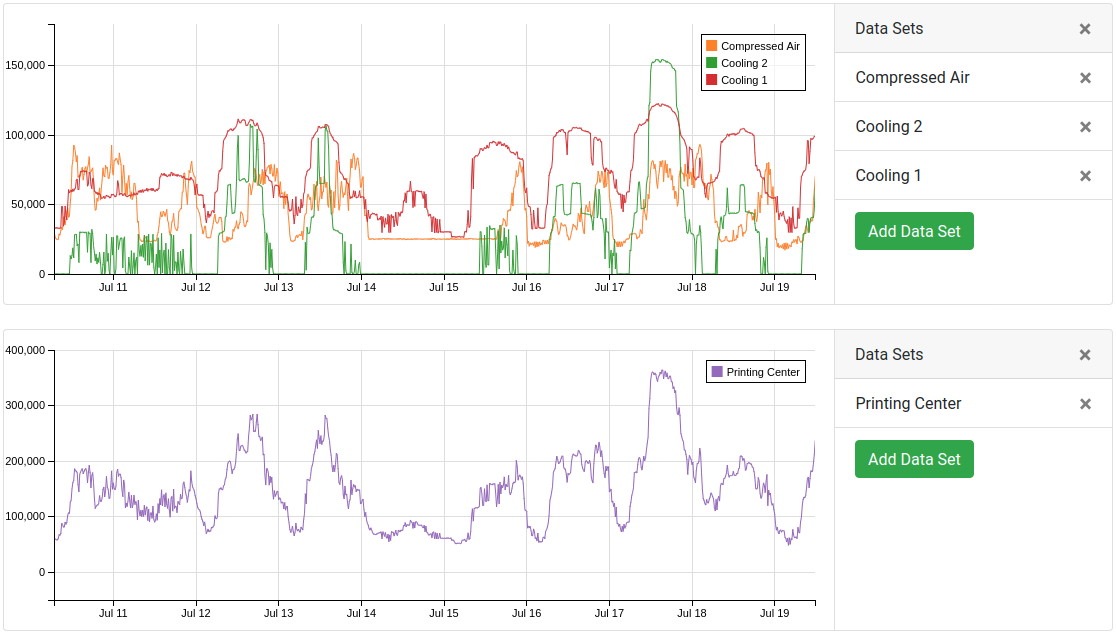}%
	\caption{Screenshot showing the graphical correlation of power consumption using the Titan Control Center.}
	\label{fig:correlation}%
\end{figure*}

The Titan Control Center provides different features for correlating power consumption data. One of these features is graphical correlation of power consumption of different machines or machine groups. Our visualization component (see Section~\ref{sec:VisualizationImpl}) provides a tool, which allows a user to compare the power consumption of multiple consumers in time series plots (see Figure~\ref{fig:correlation}). It displays multiple time series plots below each other, each containing multiple time series. The user can zoom into the plots and shift the displayed time interval. All charts are synchronized by the time domain, thus zooming or shifting one plot also effects the others \cite{Johanson2016}. This tool allows operators to analyze interesting points in time (such as outtakes or load peaks) in more detail.

Together with the newspaper printing company, we implemented a first proof of concept for correlating real-time production data with power consumption data. We correlated the printing machines' power consumption with their printing speed. For this purpose, we integrated the production management system using the Titan Flow Engine and visualized both types of data in our visualization component.
Even though we were able to show the feasibility of such a real-time correlation, we identified that for in-depth analyses, power consumption data with higher accuracy is required.
Similarly, we prototypically correlated the power consumption of air conditioning systems with weather data. We identified a high impact of the outside temperature on the power consumed for cooling and, thus, use weather data as a feature for our forecasting implementations (see Section~\ref{sec:ForecastingImpl}).

\subsection{Anomaly Detection}

The Titan Control Center envisages individual microservices for independent anomaly detection tasks and, hence, allows to choose an appropriate technique for each task. This includes individual techniques for different production environments and even for different machines.

With our pilot implementation, we already provide an Anomaly Detection microservice, which detects anomalies based on summary statistics of the previous power consumption. These statistics (e.g., per hour of week) are continuously recomputed by the Stats microservice (see Section~\ref{sec:TemporalAggregationImpl}) for each machine and machine group and published via the Control Center's messaging system. Our Anomaly Detection microservice subscribes to this statistic data stream and joins it with the stream of measurements (from real machines or aggregated groups of machines). Ultimately, this means each incoming measurement is compared to the most recent summary statistics of the corresponding point in time and machine. If the measured power consumption deviates to much from the average consumption of the respective hour and weekday, it is considered as an anomaly.
More precisely, for a measurement $x$ and summary statistics providing the arithmetic mean $\mu$ and standard deviation $\sigma$, the service computes the absolute distance from the arithmetic mean $d = \lvert x - \mu \rvert $ and tests if $d < k\sigma$, where $k$ is the configurable number of standard deviations.
All detected anomalies are again published to a dedicated data stream via the messaging system, allowing other microservices to access detected anomalies. Moreover, the microservice stores all detected anomalies in a Cassandra database.

The currently implemented method for detecting anomalies is rather simple. It does not require complex model training or manual modeling, but is not able to consider trends, seasonality over larger time periods, or external variables. We are working on extending our pilot implementation, in order to join the measurement stream with the data stream published by the forecasting service (see Section~\ref{sec:ForecastingImpl}). This implementation will consider measurements as anomalies if they deviate too much from the prediction, which is a common approach for anomaly detection (see Section~\ref{sec:AnomalyDetection})

\subsection{Forecasting}\label{sec:ForecastingImpl}

Similar to anomaly detection, we envisage individual Forecast microservices for different types of forecasts, for example, used for different power consumers.
Forecasting benefits notably from the microservice pattern since technologies used for forecasting often differ from the ones used for implementing web systems. The Titan Control Center supports arbitrary Forecast microservices, each using its own technology stack. The only requirement for a Forecast service is that it is able to communicate with other services via the messaging system.

Our pilot implementation already features a microservice that performs forecasts using an artificial neural network with TensorFlow \cite{Abadi2016}.
This neural network is trained offline using historical data and mounted into the microservice at start-up. During operation, the Forecast microservice subscribes to the stream of measurements (again monitored or aggregated) and feeds each incoming measurement into the neural network. The forecast results are stored in an OpenTSDB \cite{OpenTSDB} time series database and published to a dedicated stream via the messaging system.

In a first proof of concept, we build and trained such neural networks together with the newspaper printing company. We selected a set of machines in the company with different power consumption patterns and trained individual networks per machine. These neural networks use not only the historical power consumption of their machines as input, but also the power consumption of other machines as well as environmental data, such as the outside temperature. We deploy individual instances of our Forecast microservice for each neural network, allowing for individual forecasts of each machine. %

\subsection{Visualization}\label{sec:VisualizationImpl}

As suggested in Section~\ref{sec:Visualization}, the Titan Control Center features web applications for visualizing power consumption data. Since visualization serves as a measure to integrate the results of other measures, we also regard the visualization software components as integration of the individual analysis microservices. The Titan Control Center provides two single-page applications for visualization: a graphical user interface, tailored to the specific functions of the Titan Control Centers, and a dashboard for simple, but highly adjustable data visualizations. In the following, we describe both applications and their corresponding use cases.

\paragraph{Control Center}

The Titan Control Center user interface\footnote{\url{http://samoa.se.informatik.uni-kiel.de:8185}} serves to provide a consistent access to all functionalities of the Titan Control Center. This includes visualizing the analysis results of microservices, but also control functions for configuring microservices.
The user interface is implemented with Vue.js \cite{You2019} and D$^3$ \cite{Bostock2011}.

\begin{figure*}[t]
	\centering%
	\includegraphics[width=\linewidth]{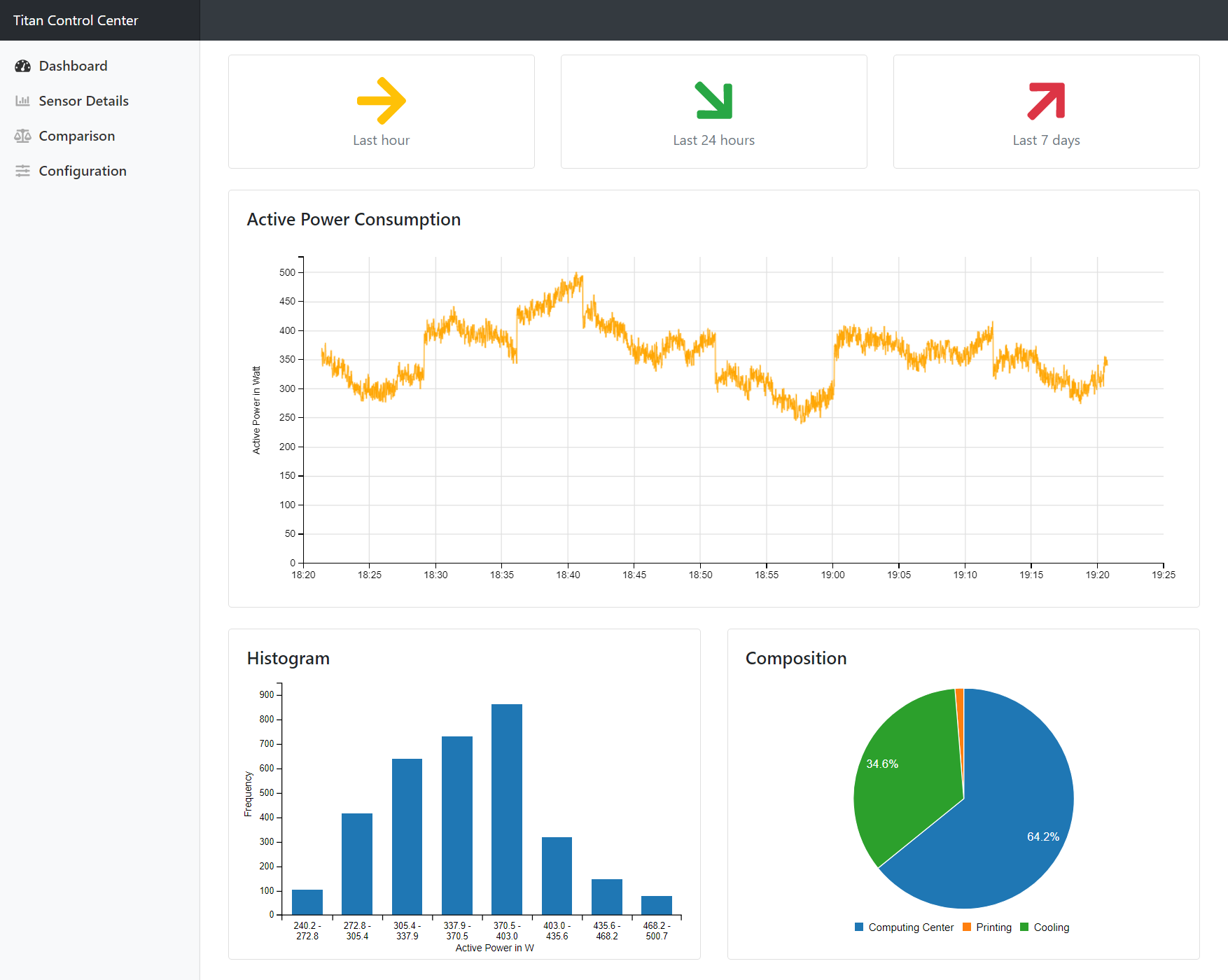}%
	\caption{Screenshot of the Titan Control Center.}
	\label{fig:visualization}%
\end{figure*}%

Figure~\ref{fig:visualization} shows a screenshot of the Titan Control Center's summary view. It consists of several components which collect and show the individual analysis results for the entire production.
A time series chart displays the power consumption in course of time. This chart is interactive by %
allowing to zoom and shift the displayed time interval.
Colored arrows indicate how the power consumption evolved within the last hour, the last 24~hours, and the last 7~days.
A histogram shows a frequency distribution of metered values serving to detect potential for load peak reduction.
A pie chart breaks down the total power consumption into subconsumers.
Line charts display the average course of power consumption over the week or the day, as provided by the Statistics microservice (see Section~\ref{sec:TemporalAggregationImpl}).
The visualizations are periodically updated with new data.
This causes, for example, the time series diagram to shift forward continuously and the arrows to change color and direction.

Apart form this summary view, our pilot implementation also provides the described types of visualization for individual machines and groups of machines. Starting from an overview of the total power consumption, a user can thus navigate through the hierarchy of all consumers.
Furthermore, the single-page application allows to graphically correlate data (see Section~\ref{sec:CorrelationImpl}) and to configure machines and machines groups maintained by the Sensor Management service.
Visualizations of forecasts and detected anomalies are currently under development.

\paragraph{Dashboard}

\begin{figure*}[t]
	\centering%
	\includegraphics[width=\linewidth]{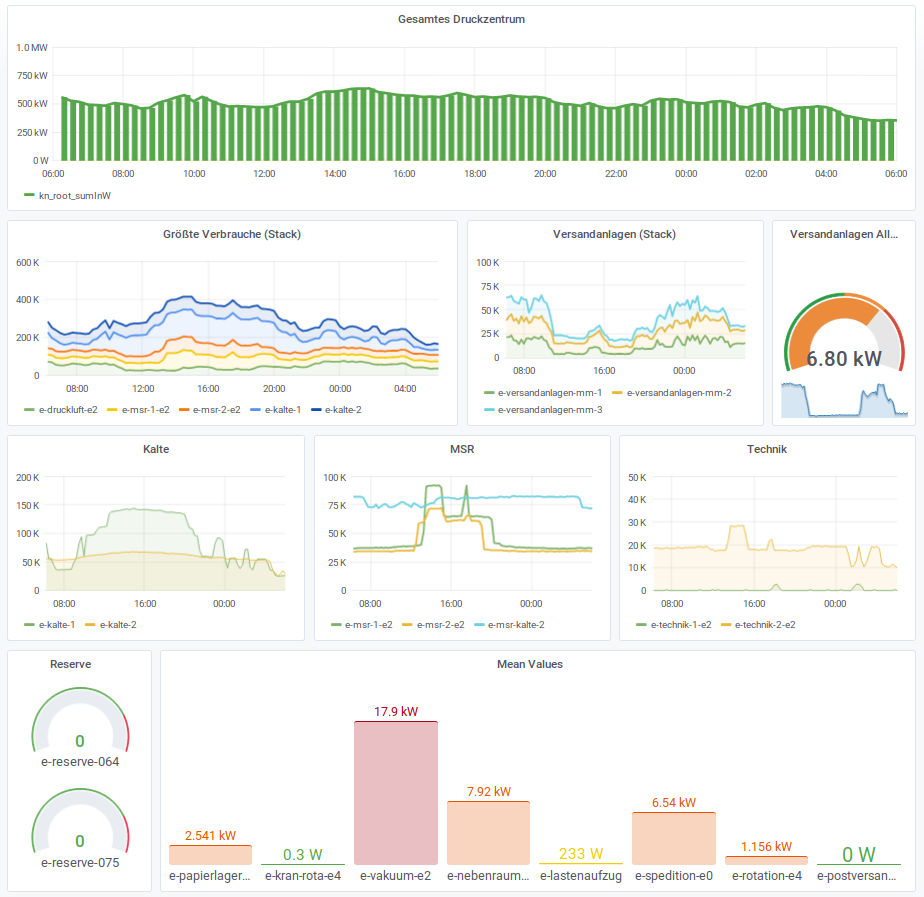}%
	\caption{Screenshot of the Titan dashboard implemented with Grafana \cite{Wetzel2019}.}
	\label{fig:grafana}%
\end{figure*}%

The second application is a pure visualization dashboard implemented with Grafana \cite{Grafana}
(see Figure~\ref{fig:grafana}). It provides a set of common visualizations such as line charts, bar charts, and gauges. As presented in Figure~\ref{fig:grafana}, we mainly display time series charts as bar or line charts.
The dashboard is highly adjustable, meaning that users can add, modify, and rearrange chart components. Such adjustments can be performed graphically and only require usage of provided interfaces. Thus, especially IT savvy production operators can customize dashboards. Moreover, they can create own dashboards and share them among users. In this way individual dashboards, for example, for management and production operators can be implemented.

In contrast to the Control Center, this dashboard does not provide any control functions (e.g., for sensor configuration) and no complex interactive visualizations (e.g., the comparison tool). 
Thus, it only serves as an extension to the Control Center, allowing for visual analysis and reporting. 
In particular, this dashboard covers use cases, where power consumption data should be integrated in existing dashboards (as it is the case in one studied enterprise) or if dashboards should be customized by production operators.

\subsection{Alerting}

Altering in the Titan Control Center is implemented using the Titan Flow Engine in the integration component. All messages that are published to the messaging system can again be consumed by the Titan Flow Engine and processed in flows. This way, production operators can create and adjust alerting flows directly within the production environment. Our pilot implementation already provides a flow that sends an email whenever an anomaly in power consumption is reported. In dedicated bricks, the operator can filter the types of anomaly an alert should be generated for and configure how the email should be sent (e.g., message and receiver).
The flow engine allows to model flows that perform arbitrary actions in the production environment when alerts are received. This includes communications with machines again, for example, to show alerts on machine monitors.

\section{Related Work}\label{sec:Related}

Analyzing industrial energy data is an emerging field of research. In the first part of this paper, we already discuss goals and measures for analyzing power consumption data, suggested in literature. A lot of research exists, in particular, on how energy data analysis can contribute to reducing the energy usage in manufacturing.
For example, \citet{Vikhorev2013} point out that making energy data available for production operators promotes energy awareness. \citet{Cagno2013} show that a lack of energy consumption information prevents implementation of energy-saving measures.
Detailed information is especially required at process and machine level for optimizing energy consumption, as highlighted by \citet{Thollander2015}.
For example, \citet{Shrouf2014b} optimize the production scheduling of a single machine for minimizing overall energy consumption costs.
For systematic monitoring and optimizing energy consumption, enterprises are moving towards establishing an energy management \cite{Cooremans2019}. \citet{Schulze2016} identify organizational measures for implementing an energy management in industry.
We extend the scope of possible goals for analyzing power consumption data by distinguishing reporting, optimization, fault detection, and predictive maintenance.

Increasing availability of smart meters and Internet of Things (IoT) adoption in the manufacturing industry enable enterprises to collect energy data in great detail.
\citet{TeschdaSilva2020} present a systematic literature review on energy management in Industry 4.0. The authors outline methods for improving energy efficiency and point out current limitations for their implementation.

\citet{Shrouf2015} highlight several benefits of IoT adoption for energy data obtained from reviewing literature and information published by European manufacturing enterprises. Their study focuses mainly on optimizing energy usage to reduce costs and improve reputation, for example, by reducing energy wastage and improving production scheduling. The authors also identify the potential of IoT energy data for predictive maintenance. We recognize large similarities to the goals identified in our pilot cases, which suggests that these goals also apply to other manufacturing enterprises.

\citet{Mohamed2019} report on opportunities provided by IoT energy data for improving energy efficiency and reducing energy costs. They focus on machine optimization as well as fault detection and predictive maintenance on individual machines. Additionally, the authors highlight how energy data can be used to improve process scheduling and building management. Moreover, they quantify saving potential in a benefit analysis. We extend this work by suggesting and implementing measures for turning identified saving potential into reality.

In Section~\ref{sec:Measures}, we describe that research suggest a variety of measures to implement the identified goals. Software systems for implementing such measures are presented, for example, by \citet{Sequeira2014} and  \citet{Rackow2015}. \citet{Yang2020} propose such a system for accessing power consumption at a university campus. However, these systems only focus on a subset of our proposed measures and are build upon different architecture decisions.

A couple of software architectures for implementing energy data analysis are suggested.
Most of these architectures \cite{Sequeira2014, Shrouf2017, Herman2018, Liu2018} follow the Lambda architecture pattern \cite{Marz2015}. Such architectures deploy a speed layer for fast online processing and a batch layer for correct offline processing of data. We pursue a more recent architectural style of processing data exclusively online (also referred to as Kappa architecture) \cite{Kreps2014} by utilizing Apache Kafka's capabilities for reprocessing distributed, replicated logs \cite{Wang2015}.
Additionally, we combine this with the microservice architecture pattern and design dedicated, encapsulated microservices per analytics task.
Benefits of using microservices and, in particular, the associated concept of polyglot persistence for analyzing industrial energy usage are highlighted by \citet{Herman2018} and \citet{Henning2019a}.

Big data analytics of energy consumption heavily relies on cloud computing \cite{Shrouf2014a, Herman2018,Sequeira2014,Mohamed2018,Yang2020}. \citet{Sequeira2014} propose \textit{cloud connector} software components for integrating data from energy meters. Recent studies suggest to apply fog computing for integrating production data in general \cite{Qi2019} and energy consumption data in particular \cite{Mohamed2019}. 
We follow the suggestions of \citet{Pfandzelter2019} to deploy data analytics using stream processing in the cloud and data preprocessing and event processing in the fog.
\citet{Szydlo2017} present how data transformation at fog computing nodes can be implemented using flow-based programming and graphical dataflow modeling.

\section{Conclusions and Future Work}\label{sec:Conclusions}

Analyzing power consumption data in manufacturing enterprises promises to achieve goals of categories such as reporting, optimization, fault detection, and predictive maintenance. Based on reviewed literature and expert interviews, we suggest to implement the following measures in software for achieving these goals: real-time data processing, multi-level monitoring, temporal aggregation, correlation, anomaly detection, forecasting, visualization, and alerting.
Accompanied by real goals in two manufacturing enterprises, we show how microservices, stream processing, and fog computing can serve for implementing the proposed measures in a power consumption analytics platform.

For future work, we plan to take advantage of the modular architecture of the Titan Control Center by extending our pilot implementations. In particular, ongoing research focuses on developing more precise forecast and anomaly detection approaches as well as detailed visualizations. Further, we plan to conduct extensive evaluations in our studied enterprises.

\section*{Acknowledgments}
This research is funded by the German Federal Ministry of Education and Research (BMBF) under grand no.\ 01IS17084 and is part of the Titan project (\url{https://www.industrial-devops.org}).

\section*{References}
\bibliography{references.bib}

\end{document}